\documentclass[12pt]{article}
\usepackage[dvips,usenames]{color}
\usepackage[dvips]{graphicx}
\usepackage[letterpaper,dvips,body={8in,11in},vmargin=2cm,hmargin=2cm]{geometry}
\usepackage{xspace}
\usepackage[american]{babel}
\usepackage[latin1]{inputenc}
\usepackage[T1]{fontenc}
\usepackage{ae}
\usepackage{aecompl}
\usepackage[dvipdfm,
            bookmarks     = true,
            breaklinks    = true,
            hypertexnames = false,
            hyperfigures  = true,
            colorlinks    = true,
            urlcolor      = blue]
{hyperref}
\hypersetup{
  pdftitle    = {Daniel's Draft - v3},
  pdfauthor   = {Daniel Doro Ferrante <danieldf@olympus.het.brown.edu>},
  pdfsubject  = {High Energy Physics - Theory},
  pdfcreator  = {LaTeX2e with hyperref package},
  pdfproducer = {dvipdfm},
  pdfkeywords = {Quantum Field Theory, Lattice, Quantum Gravity, Symmetry Breaking,
                 Topology, Index Theorems},
  pdfview     = {FitH}
}
\usepackage{fancyhdr}
\fancypagestyle{plain}{%
\fancyhf{} 
\fancyfoot[C]{\bfseries\thepage} 

}

\newcommand{\helv}{%
\fontfamily{phv}\fontseries{b}\fontsize{9}{11}\selectfont}
\usepackage{curves}
\usepackage{boxedminipage}
\usepackage[nice,tight]{units}
\usepackage{amsmath,amsfonts,amstext,amssymb,amsbsy,amsopn,amsthm,eucal}
\DeclareMathAlphabet{\mathpzc}{T1}{pzc}{m}{it} 
\usepackage{mathrsfs}
\usepackage{wasysym}
\usepackage{dsfont}
\makeatletter
\def\equalsfill{$\m@th\mathord=\mkern-7mu
  \cleaders\hbox{$\!\mathord=\!$}\hfill
  \mkern-7mu\mathord=$}
\makeatother
\def\hksqrt{\mathpalette\DHLhksqrt}
\def\DHLhksqrt#1#2{\setbox0=\hbox{$#1\sqrt{#2\,}$}\dimen0=\ht0
  \advance\dimen0-0.2\ht0
  \setbox2=\hbox{\vrule height\ht0 depth -\dimen0}%
{\box0\lower0.4pt\box2}}

\DeclareMathOperator{\tr}{tr}
\DeclareMathOperator{\ed}{d\!}
\DeclareMathOperator{\re}{\mathsf{Re}}
\DeclareMathOperator{\arcsinh}{arcsinh}

\newcommand{\ie}{\textit{i.e.}\xspace}
\newcommand{\g}{\ensuremath{\boldsymbol{g}}\xspace}
\newcommand{\gE}{\ensuremath{\boldsymbol{\tilde{g}}_{E}}\xspace}

\newcommand{\hm}[2]{\ensuremath{\langle#1,#2\rangle}\xspace}

\begin{document}
\begin{titlepage}
  \begin{flushright}
    {BROWN-HET-1455} \\
    \texttt{hep-th/0609190}
  \end{flushright}

  \bigskip
  \begin{center}
    {\LARGE \textbf{\textsf{From Symmetry Breaking to Topology Change I}}} \\
    \vspace{1cm}
    \Large{\textsf{Daniel D. Ferrante and Gerald S. Guralnik}} \\
    \bigskip
    \textsf{Physics Department\footnote{\texttt{WWW:}
        \href{http://chep.het.brown.edu/}{\texttt{chep.het.brown.edu}}}} \\
    \textsf{Brown University, Providence --- RI. 02912}
    \date{\today}
  \end{center}

  \bigskip
  \begin{abstract}
    \noindent By means of an analogy with Classical Mechanics and Geometrical
    Optics, we are able to reduce Lagrangians to a kinetic term only. This form
    enables us to examine the extended solution set of field theories by finding
    the geodesics of this kinetic term's metric. This new geometrical standpoint
    sheds light on some foundational issues of QFT and brings to the forefront
    core aspects of field theory.
  \end{abstract}

  \tableofcontents
\end{titlepage}
\section{Introduction}\label{sec:intro}
Classical gauge theories are well described through differential geometry, where
a gauge field is represented by a connection on a principal fibre bundle $P$ for
which the structure group is the symmetry group of the theory; see, for example,
\cite{frankel,nakahara,darling,ladies,kerbrat,fulpnorris,mayer,sardanashvily,simpson}
and references therein.

The phenomenon of spontaneous symmetry breaking also has its own geometric
formulation \cite{ladies,kerbrat,fulpnorris,mayer,sardanashvily,simpson,neeman}: the
reduction of the principal bundle $P$. To this classical picture, the question
that arises is that of how does a quantum field theory select its vacua given by
the different possible reductions of the principal bundle $P$. One canonical
answer to this question is given by the so-called ``Spontaneous Symmetry
Breaking'' mechanism, \ie, radiative corrections (beyond the tree-level
approximation) cause the QFT to ``jump'' from the symmetric phase to the
broken-symmetric one(s) \cite{weinbergcoleman}.

However, \cite{weinbergcoleman} already expresses concerns about
some issues (e.g., the last paragraph of page 1894 and its continuation on page
1895) that are more explicitly treated in \cite{garciaguralnik}. (For more
examples about these issues, see \cite{borcherds} --- pages 27 and 28 discuss
the asymptotic nature of the series expansion, their maximum accuracy and issues
regarding Borel summation --- and \cite{fredenhagen}.)

The analysis of the boundary conditions of the Schwinger-Dyson equation being
responsible for the different phases of the theory done in \cite{garciaguralnik}
is more in the spirit of the self-adjoint extension of the associated
Hamiltonian, as done in \cite{reedsimon,dunfordschwartz}. Although it may not
seem so at first, \cite{garciaguralnik} is equivalent to a latticized approach to
the QFT and, as such, requires 2 types of infinite limits in order to give rise
to its continuum version: the thermodynamic (number of particles in the box) and
the volume (size of the box) limit.

These limits do not commute and fiddling with them (as done in
\cite{garciaguralnik}) is analogous to resumming the perturbative series (as
done in \cite{weinbergcoleman}).

The present work has the goal of developing a geometrical method that brings to
the foreground issues of phase structure in QFT and clearly showing that these
different phases are topologically inequivalent. The fact, that the different
configuration spaces for the distinct solutions of the equation of motion of a
given Lagrangian have different topologies, shows that expansions must be
performed separately for each solution of a theory, \ie, each phase has to be
regarded and treated as a separate theory.

In this sense, this work streamlines how the parameters of a theory (mass,
coupling constants, \textit{etc.}) determine the topology of each phase which,
combined with the picture presented in \cite{garciaguralnik}, gives a
prescription of how to examine the solution space of a field theory: The
boundary conditions of the Schwinger-Dyson equations determine the parameters of
the theory which, in turn, determine its phase (and topology). To this we can
add the Renormalization Group framework and see how the phases of a theory (and
their topologies) change: as one flows the Renormalization Group, the theory
moves through its phases (and its inequivalent topologies). Note that just like
in \cite{garciaguralnik}, the number of different solutions, found in this work,
to a theory whose potential energy is given by a polynomial of degree $n$ in the
fields (with no derivative couplings) is $n-1$ times infinity, \ie, $n-1$
solutions per spacetime point.

Moreover, when applied to String Theory, under the Landscape paradigm, this
result explicitly says that each different minima of the Landscape potential has
a different spacetime topology. Therefore, spontaneous symmetry breaking would
lead to topological change in spacetime, a key ingredient in String Theory ---
doing away with the use of a Coleman-De Luccia \cite{colemandeluccia} type of
mechanism.

In section \ref{sec:jmqft} we will put the method forward and demonstrate it on
the free field. In section \ref{sec:apps} we will apply this method for 3
different theories: $\lambda\, \phi^4$, Landau-Ginzburg and
Seiberg-Witten. (Note that the main difference between $\lambda\, \phi^4$ and
Landau-Ginzburg and Seiberg-Witten is that the first one is a scalar field
theory without gauge symmetry, while the latter two do have gauge fields.) Then,
in section \ref{sec:cc}, we will conclude summarizing our results.
\section{Quantum Field Theory and the Jacobi Metric} \label{sec:jmqft}
A typical action for a scalar field in QFT has the form,
\begin{align*}
  S[\phi] &= \int K(\pi,\pi) - V_p(\phi) \, d^{n}x \; ;\\
  K(\pi,\pi) &= \frac{1}{2}\, \g(\pi,\pi) = \frac{1}{2}\, g_{\mu\,\nu}\,
    \pi^{\mu}\,\pi^{\nu} \; .
\end{align*}
where $K$ is the kinetic quadratic form (bilinear and, in case $\pi\in\mathbb{C}$,
hermitian; defining the inner product \g) and $V_p$ is the potential, the index $p$
denoting coupling constants, mass terms, \textit{etc}\/.

We want to reparameterize our field using the arc-length parameterization such
that $\boldsymbol{\tilde{\g}}(\tilde{\pi},\tilde{\pi}) = \tilde{g}_{\mu\,
  \nu}\tilde{\pi}^{\mu}\, \tilde{\pi}^{\nu} = 1$, where
$\boldsymbol{\tilde{\g}}$ is the new metric and $\tilde{\pi}^{\mu}$ is the
momentum field redefined in terms of $\boldsymbol{\tilde{\g}}$ (assuming
$V_p(\phi)$ contains no derivative couplings). That is, we want to dilate or
shrink our coordinate system in order to obtain a conformal transformation of
the metric that normalizes our momentum field.

In order to do that, we will borrow an idea from classical mechanics called
\emph{Jacobi's metric} \cite{drg}. In the Newtonian setting, Jacobi's metric
gives an intrinsic geometry for the configuration space, where dynamical orbits
become geodesics (\ie, it maps every Hamiltonian flow into a geodesic one;
therefore, solving the equations of motion implies finding the geodesics of
Jacobi's metric and \textit{vice-versa}). This can be done for any closed
non-dissipative system (with total energy $E$), regardless of the number of
degrees of freedom.

Before we proceed any further, let us take a look at a simple example in order
to motivate our coming definition of Jacobi's metric. Let us consider a model
similar to the harmonic oscillator (without further considerations, we simply
allow for the analytic continuation of the frequency, $\omega^2 = \mu$ or
$\omega^2 = i\,\mu$) in Quantum Mechanics (1-dimensional QFT). Its
Lagrangian is given by $L = \tfrac{1}{2}\, \dot{q}^2 \mp \tfrac{\mu}{2}\, q^2$
($\mu > 0$), from which we conclude that, for a fixed $E$ such that $E =
\tfrac{1}{2}\, (\dot{q}^2 \pm \mu\,q^2)$, we have $\dot{q} = \tfrac{dq}{dt} =
\hksqrt{2\, (E \mp \tfrac{\mu}{2}\, q^2)}$.

So, in the spirit of what was said above, we want to find a reparameterization
of the time coordinate in order to have the normalization $\dot{q} = 1$, where
the dot represents a derivative with respect to this new time variable.

Thus,

\begin{align*}
  \frac{dq}{dt} &= \frac{dq}{ds}\, \frac{ds}{dt} = \hksqrt{2\, \biggl(E \mp
    \frac{\mu}{2}\, q^2\biggr)} \; \\
  \text{if}\quad \frac{ds}{dt} &= \hksqrt{2\, \biggl(E \mp \frac{\mu}{2}\,
    q^2\biggr)} \Rightarrow \frac{dq}{ds} \equiv 1\;. \\
  \therefore\; ds &= \hksqrt{2\, \biggl(E \mp \frac{\mu}{2}\, q^2\biggr)}\, dt\; ; \\
  \Rightarrow\; \gE &= 2\, \biggl(E \mp \frac{\mu}{2}\, q^2\biggr)\, \g \; .
\end{align*}

As expected, the new metric, \gE, is a conformal transformation of the original
one, \g; and under this new metric, our original Lagrangian is simply written
as,

\begin{align*}
  L(q,\dot{q}) &= \frac{1}{2}\, \dot{q}^2 \mp \frac{\mu}{2}\, q^2 \; ;\\
  &= \frac{1}{2}\, g_{i\, j} \dot{q}^i\,\dot{q}^j \mp\frac{\mu}{2}\, q^2\; ;\\
  &= \frac{1}{2}\, \g(\dot{q},\dot{q}) \mp\frac{\mu}{2}\, q^2\; ;\\
  \therefore\; L(q,\dot{q}) &= \gE(\dot{q},\dot{q}) \; .
\end{align*}

Now, as we said above \cite{drg}, the Euler-Lagrange equations for this
conformally transformed Lagrangian are simply the geodesics of the metric
\gE. The task of finding the geodesics $\gamma(s)$ of the \gE metric can be more
easily accomplished with the help of the normalization condition,
$\gE\bigl(\tfrac{d\gamma}{ds},\tfrac{d\gamma}{ds}\bigr) = 1$, and the initial
condition $\gamma(s=0) = 0$, given that, in this fashion, we only need to solve
a first order differential equation:

\begin{align*}
  \gE\biggl(\frac{d\gamma}{ds}, \frac{d\gamma}{ds}\biggr) &= 1 \; ;\\
  \Rightarrow\; 2\, \biggl(E \mp \frac{\mu}{2}\, \gamma^2\biggr)\,
    \g\biggl(\frac{d\gamma}{ds}, \frac{d\gamma}{ds}\biggr) &= 1 \; ;\\
  \therefore\; \frac{d\gamma}{ds} &= \hksqrt{\frac{1}{2\,\bigl(E \mp \frac{\mu}{2}\,
      \gamma^2\bigr)}} \; ;\\
  \text{with the initial condition}\quad \gamma(0) &= 0 \; .
\end{align*}

Analytically solving the equation above yields 2 possible answers, depending on
the particular form of the potential (note that $\mu > 0$ in both cases):

\begin{enumerate}
\item $V_{+} = +\mu\, \gamma^{2}/2$: For the case of a positive pre-factor, we
  get that $\gamma\, \hksqrt{\mu\, (2\, E - \mu\, \gamma^{2})} + 2\, E\,
  \arcsin\bigl(\gamma\, \hksqrt{\mu/2E}\bigr) - 2\, s\, \hksqrt{\mu} = 0$; and
\item $V_{-} = -\mu\, \gamma^{2}/2$:  For the case of a negative pre-factor, we
  find that $\gamma\, \hksqrt{\mu\, (2\, E + \mu\, \gamma^{2})} + 2\, E\,
  \arcsinh\bigl(\gamma\, \hksqrt{\mu/2E}\bigr) - 2\, s\, \hksqrt{\mu} = 0$.
\end{enumerate}

It is clear from the expression for $V_{+}$ that $\gamma^{2} \leqslant 2E/\mu$,
\ie, the length of the [classical] geodesic is bounded; this does not happen
with $V_{-}$.

These geodesics, $\gamma_{\pm}$, clearly depend on the parameters $E$ and $\mu$;
therefore, in order to plot $\gamma(s)$, we have to make 2 distinct choices:
$E/\mu = 1$ (left plot) and $E/\mu = -1$ (right plot). The last plot
comparatively depicts both geodesics.

\vspace{.7cm}
\begin{center}
  \includegraphics[scale=0.5]{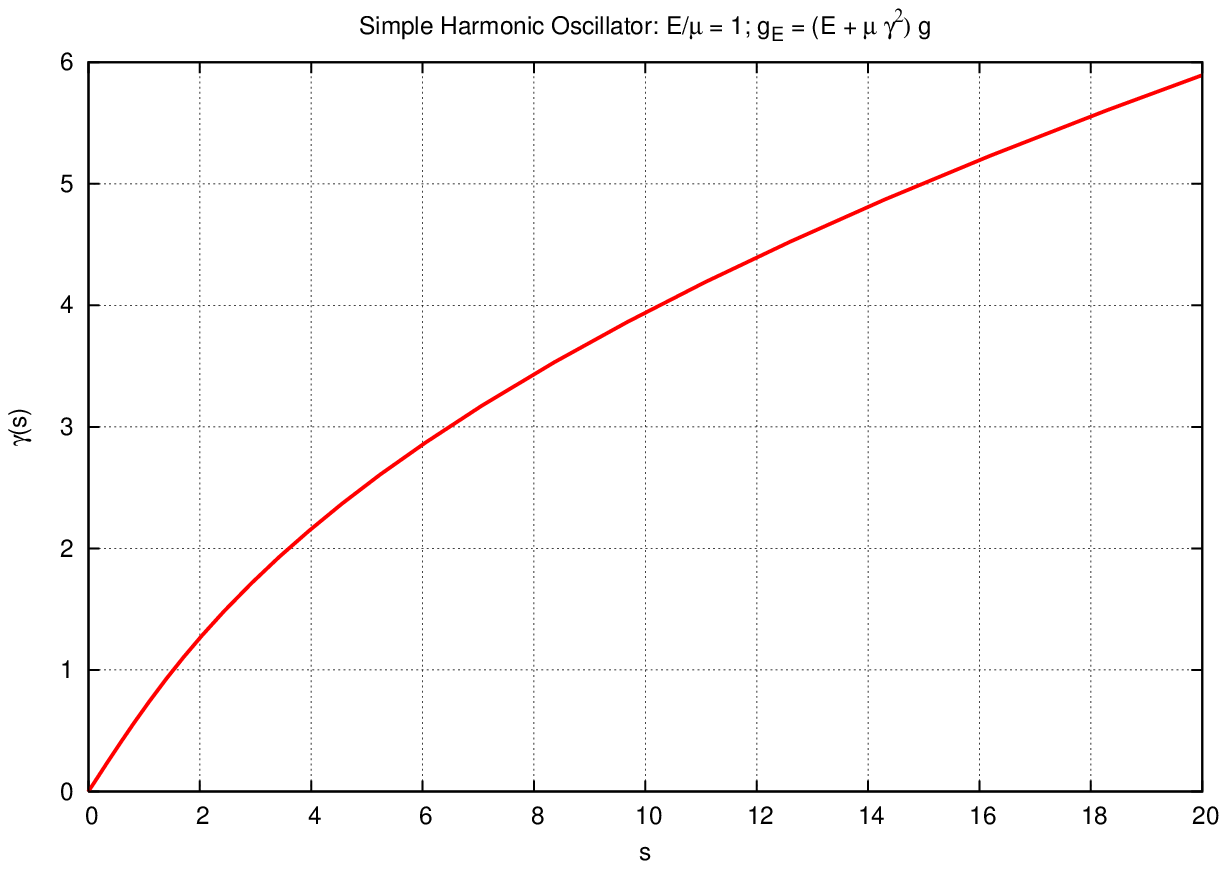} \hfill
  \includegraphics[scale=0.5]{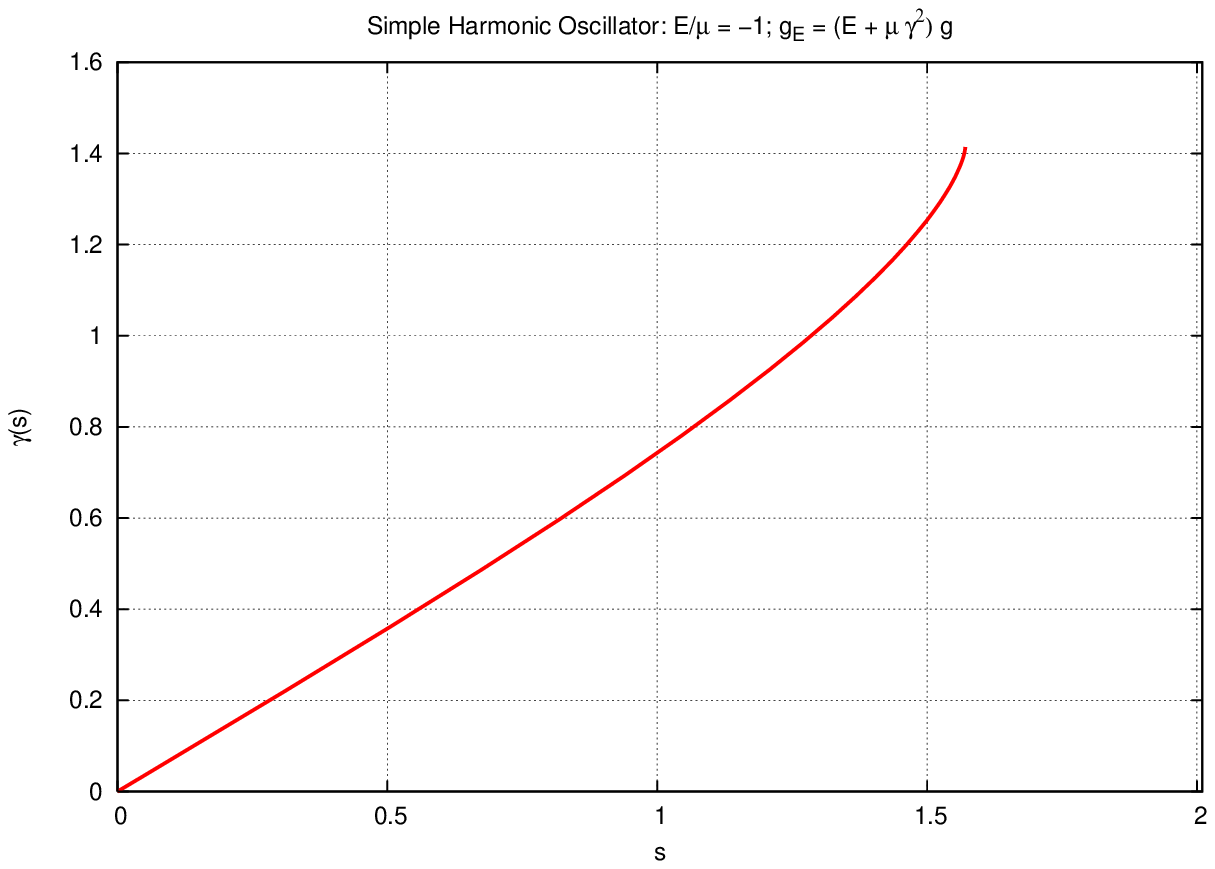} \hfill\\
  \includegraphics[scale=0.5]{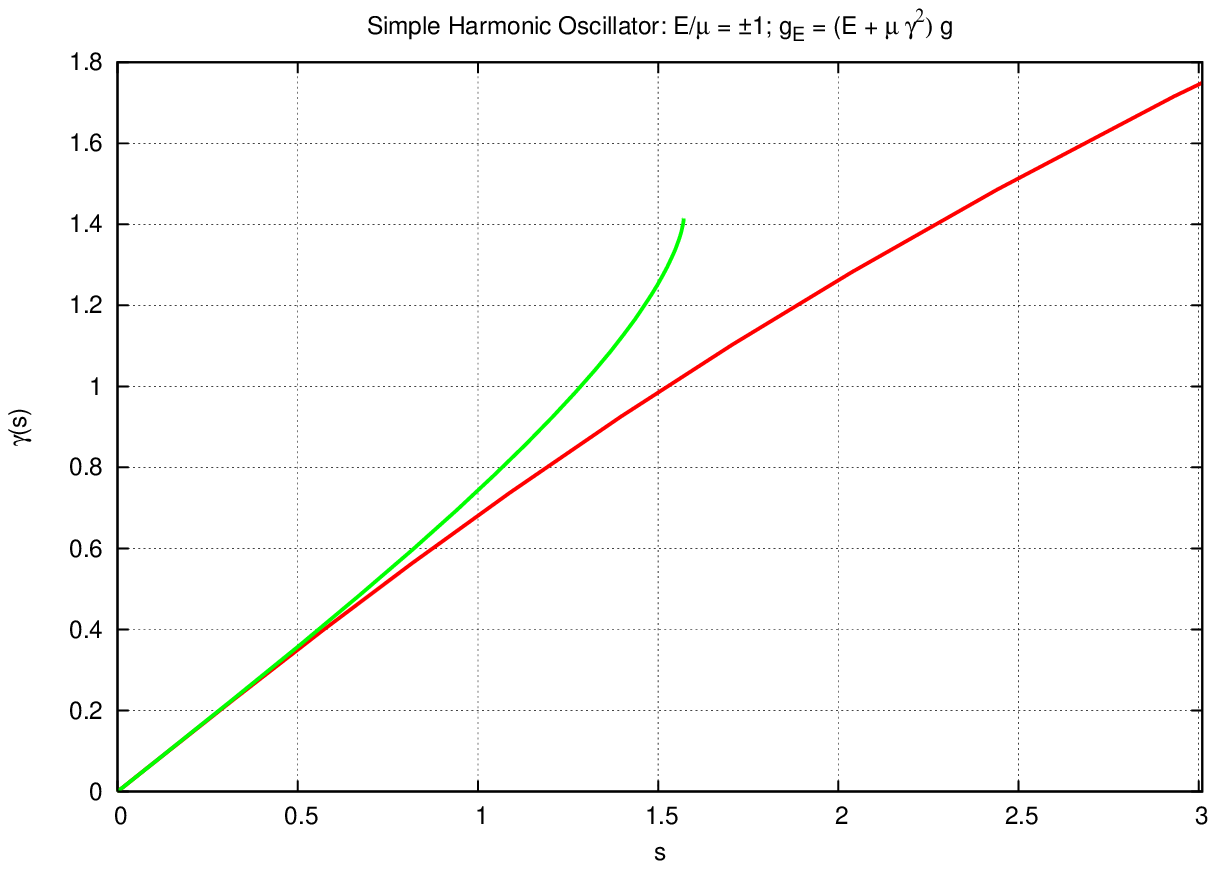}
\end{center}

This ``harmonic oscillator'' model already portrays the features that we want to
identify in forthcoming applications of this technique: changing the values of
the parameters of the potential we can identify 2 distinct types of geodesics
which will be related to different phases on the coming examples.

Note that we do not need to engage on discussions of the stability (boundedness)
of the potential for all we wanted to show was that the parameters of the
potential will give rise to qualitatively and quantitatively different geodesic
solutions. We could have used any other [bounded] potential, e.g., $V(q) =
-\tfrac{\mu}{2}\, q^2 + \tfrac{\lambda}{4}\, q^4$ and done an analogous
analysis. However, we just wanted to keep the number of parameters $(E, \mu,
\lambda)$ low, hence our choice.

At this point, the generalization of Jacobi's metric to the field theoretical
setting can be accomplished via the definition of the following conformally
transformed metric:

\begin{align}
  \nonumber
  L &= \frac{1}{2}\, \g(\pi,\pi) - V_p(\phi) \; ; \\
  \nonumber
  &\equiv \gE(\pi,\pi) \; ;\\
  \intertext{\hspace{5cm} where,}
  \label{eq:jacobimetric}
  \gE &= 2\, \bigl(E - V_p(\phi)\bigr)\, \g \; ;
\end{align}
and $E$ is the total energy of the system.
\section{Applications} \label{sec:apps}
In what follows, we will apply this technique for finding distinct geodesics to
different potentials, showing that the different geodesics pertain to different
phases of the QFT at hand.

We will see from the examples below that the presence of gauge symmetry does not
affect our results, as expected, once all we are doing is conformally changing
the metric of the phase space.

%
\subsection{The $\boldsymbol{\lambda\, \phi}^{\mathbf{4}}$ Potential}\label{subsec:lp4}
This theory is defined by the Lagrangian $L = \tfrac{1}{2}\,
\bigl(\g(\pi,\pi) - \mu\, \phi^2 - \tfrac{\lambda}{2}\,
\phi^4\bigr)$, \ie, the potential is given by $V(\phi) = \tfrac{1}{2}\,
\bigl(\mu\, \phi^2 + \tfrac{\lambda}{2}\, \phi^4\bigr)$, with reflection
symmetry $\mathbb{Z}_2$, $\phi \mapsto -\phi$.

Canonically, the reasoning is that the mass parameter $(\mu)$ has
to change sign so that the fields go from the symmetric phase to the
broken symmetric one. Mathematically speaking, this represents a change in the
solution space given by $\mathscr{F} = \ker\bigl\{(\Box + \mu +
\lambda\phi^2)\bigr\} \equiv \bigl\{ \phi \;|\; (\Box + \mu + \lambda\phi^2)\,
\phi = 0\bigr\}$, \ie, different parameters (in this case, $\mu$ and $\lambda$)
will give rise to equivalence classes of solutions: a symmetric phase
equivalence class and an equivalence class for each broken symmetric phase. This
division of the solution space into equivalence classes labelled by the
parameters of the potential is true for classic and for quantum field theories.

In this particular case, these equivalence classes are labeled by the parameter
$\tfrac{\mu}{\lambda}$ and there are, essentially, 2 of them: The symmetric
equivalence class is parameterized by $\tfrac{\mu}{\lambda} > 0$ and the
broken symmetric equivalence class is parameterized by $\tfrac{\mu}{\lambda} <
0$. Note that $\lambda \neq 0$, otherwise we would be treating the free scalar
field QFT.

However, QFTs can undergo a dynamical process, Spontaneous Symmetry Breaking,
whereby the quantum fields move from one solution space (equivalence class) to
another, namely from the symmetric phase to the broken symmetric one
\cite{weinbergcoleman} (note that this is not possible in classical field
theories, given that the origin of this symmetry breaking dynamics is due to
quantum effects).

Therefore, the solution space equivalence classes (labeled by $p =
\tfrac{\mu}{\lambda}$) change from the symmetric $\mathscr{S}_{p>0} = \ker\{\Box
+ \lambda\, \phi^2 + \mu\}$ to the broken symmetric $\mathscr{B}_{p<0} =
\ker\{\Box + \lambda\, \phi^2 - \mu\}$. Thence, their geometry changes
accordingly, a fact that can be measured by the procedure outlined above, using
$\gE = 2\, \bigr(E \pm \tfrac{\mu}{2}\, \gamma^2 + \tfrac{\lambda}{4}\,
\gamma^4\bigr)\, \g$ and solving the normalization condition $\gE(\gamma',\gamma')
= 1$, where $\gamma(s)$ is the geodesic we want to compute and $\gamma' =
\tfrac{d\gamma}{ds}$, $s$ being the arc-length parameter.

However, this naïve, albeit common, reasoning is just part of the story. The
full picture presents itself only upon a more detailed analysis of
$\int\hksqrt{E - \mu\, \gamma^{2}/2 - \lambda\, \gamma^{4}/4}\, d\gamma$, which
is the [elliptic] integral that needs to be solved in order to find
$\gamma(s)$. Therefore, it is useful to consider the polynomial $P(\gamma) =
(\gamma^{2} - r_1)\, (\gamma^{2} - r_2)$, where $(-\lambda/4)\, P(\gamma) = E -
\mu\, \gamma^{2}/2 - \lambda\, \gamma^{4}/4$ and $r_{1,2} = -\bigl(\mu \pm
\hksqrt{4\, E\, \lambda + \mu^{2}}\bigr)/\lambda$. It is the discriminant of
this polynomial $P(\gamma)$ that will, ultimately, determine the different
phases of the theory: $\Delta = (r_1 - r_2)^{2} = \lambda\, E + \mu^{2}/4
\lesseqgtr 0$.

The 2 inequalities, $\Delta > 0$ and $\Delta < 0$, are related by the analytic
continuation (and, therefore, their boundary conditions for their respective
Schwinger-Dyson equations are different) of the mass parameter, \ie,
$\mu \mapsto -\mu$; meanwhile, the case $\Delta = 0$ has to be
computed separately. This whole framework is potentially a very interesting
result, once discriminants are closely related and contain information about
ramifications (``branching out'' or ``branches coming together'') in Number
Theory. In this sense, it is more appropriate to label the different solution
space equivalence classes using $\Delta$ rather than $p = \mu/\lambda$: The
symmetric one is given by $\mathscr{S}_{\Delta > 0}$, while the broken symmetric
one is now seen to be split into 2, $\mathscr{B}_{\Delta = 0}$ and
$\mathscr{B}_{\Delta < 0}$.

The analytical answers for the geodesic $\gamma(s)$ are found to be given by the
following implicit equations:

\begin{description}
\item[$\mathbf{\Delta = 0:}$]
  \begin{equation}
    \label{eq:deltaEQ0}
    \frac{1}{3}\, \gamma^{3} + \frac{\mu}{\lambda}\, \gamma + s = 0\; ;
  \end{equation}
\item[$\mathbf{\Delta > 0:}$] 
  \begin{equation}
    \label{eq:deltaGT0}
    \begin{split}
      3\, s &+ \gamma\, \hksqrt{(\gamma^{2} - r_1)\,(\gamma^{2} - r_2)} + 2\, r_1\,
      \hksqrt{r_2}\,
      F\biggl(\frac{\gamma}{\hksqrt{r_1}}\boldsymbol{;}
      \hksqrt{\tfrac{r_1}{r_2}}\biggr) -\\
      &- \frac{2}{3}\, \frac{\mu}{\lambda}\,
      \hksqrt{r_2}\, \Biggl[F\biggl(\frac{\gamma}{\hksqrt{r_1}}\boldsymbol{;} 
      \hksqrt{\tfrac{r_1}{r_2}}\biggr) - E
      \biggl(\frac{\gamma}{\hksqrt{r_1}}\boldsymbol{;}
      \hksqrt{\tfrac{r_1}{r_2}}\biggr)\Biggr] = 0 \; ;
    \end{split}
  \end{equation}
\item[$\mathbf{\Delta < 0:}$] 
  \begin{equation}
    \label{eq:deltaLT0}
    \begin{split}
      3\, s &+ \gamma\, \hksqrt{(\gamma^{2} - r_1)\,(\gamma^{2} - r_2)} + 2\, r_1\,
      \hksqrt{r_2}\,
      F\biggl(\frac{\gamma}{\hksqrt{r_1}}\boldsymbol{;}
      \hksqrt{\tfrac{r_1}{r_2}}\biggr) -\\
      &- \frac{2}{3}\, \frac{\mu}{\lambda}\,
      \hksqrt{r_2}\, \Biggl[F\biggl(\frac{\gamma}{\hksqrt{r_1}}\boldsymbol{;} 
      \hksqrt{\tfrac{r_1}{r_2}}\biggr) - E
      \biggl(\frac{\gamma}{\hksqrt{r_1}}\boldsymbol{;}
      \hksqrt{\tfrac{r_1}{r_2}}\biggr)\Biggr] = 0 \; .
    \end{split}
  \end{equation}
\end{description}

Note that, in the equations above, $F(z;k)$ is the incomplete elliptic integral
of the first kind, while $E(z;k)$ is the incomplete elliptic integral of the
second kind. Moreover, $r_{1,2}$ are defined (as shown above) as the solutions to
the polynomial equation $P(\gamma) = 0$: $r_{1,2} = \mu \pm \hksqrt{\mu^{2} +
  4\, \lambda \, E}/(-\lambda)$. On top of this, although \eqref{eq:deltaGT0}
and \eqref{eq:deltaLT0} have the same form, they will yield distinct solutions,
once the relation given by $\Delta = \lambda\, E + \mu^{2}/4$ will either be
positive or negative, which, in turn, affects the outcome of $r_{1,2}$.

The graphical results are shown below, where $\tfrac{E}{\lambda} = 1,
\gamma(0) = 0$ and $\Delta$, respectively, assumes positive
($\Delta > 0$), null ($\Delta = 0$) and negative ($\Delta < 0$) values.

\begin{center}
  \includegraphics[scale=0.45]{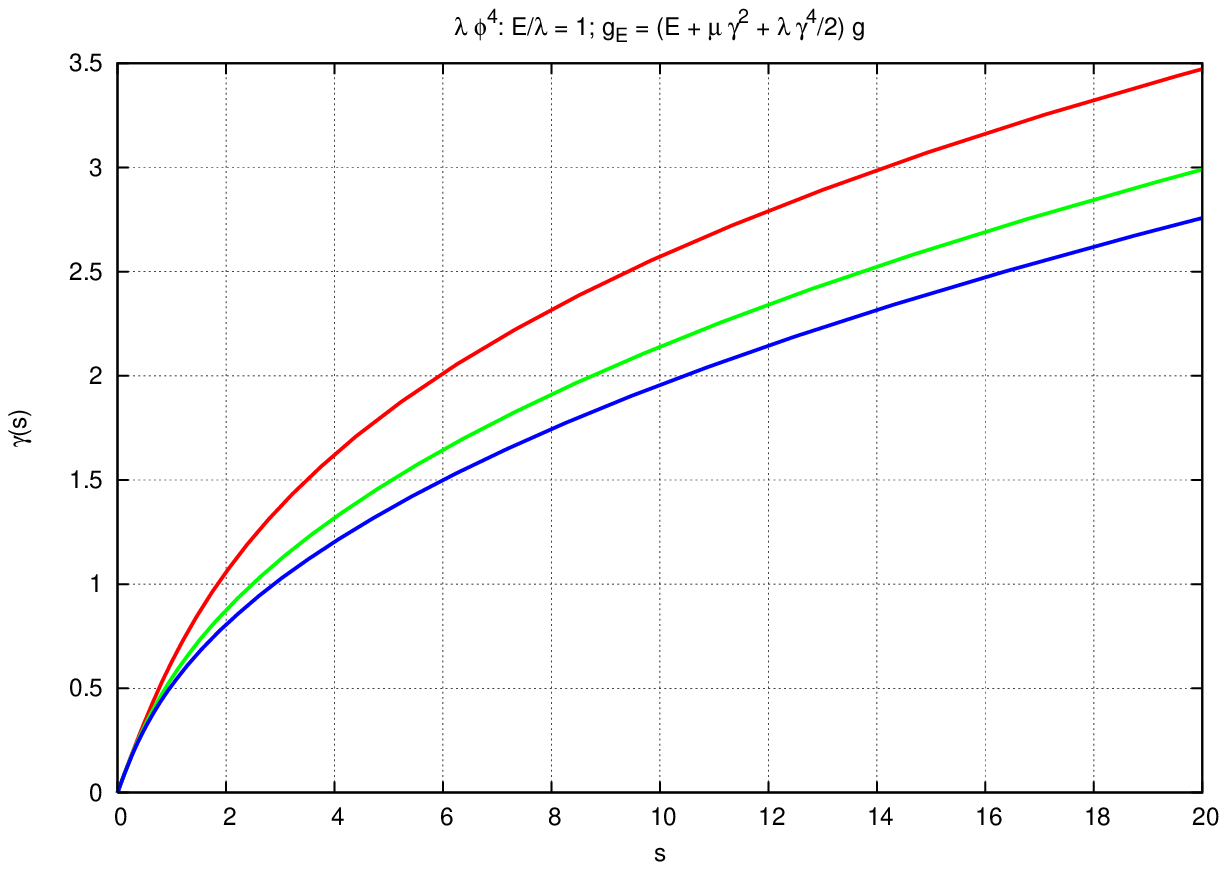} \hfill
  \includegraphics[scale=0.45]{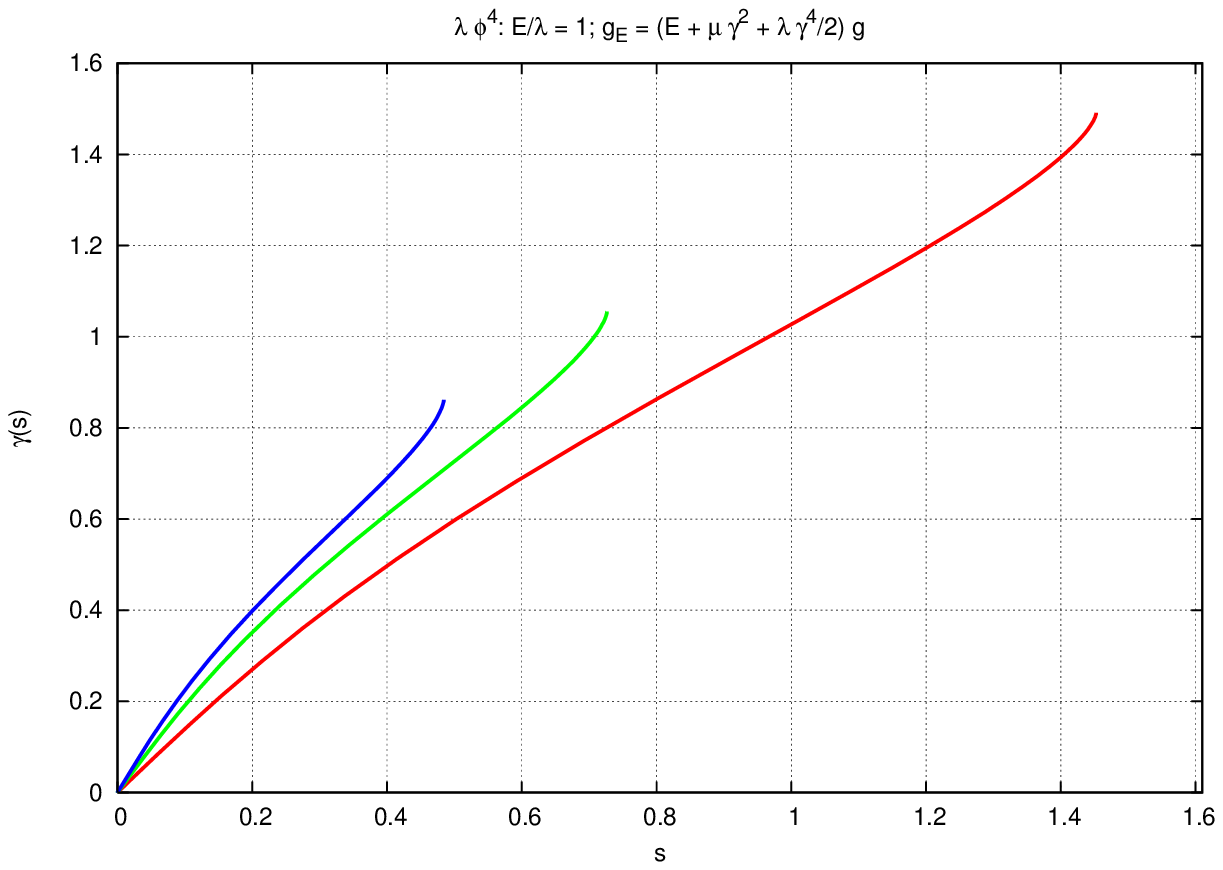}\hfill
  \includegraphics[scale=0.45]{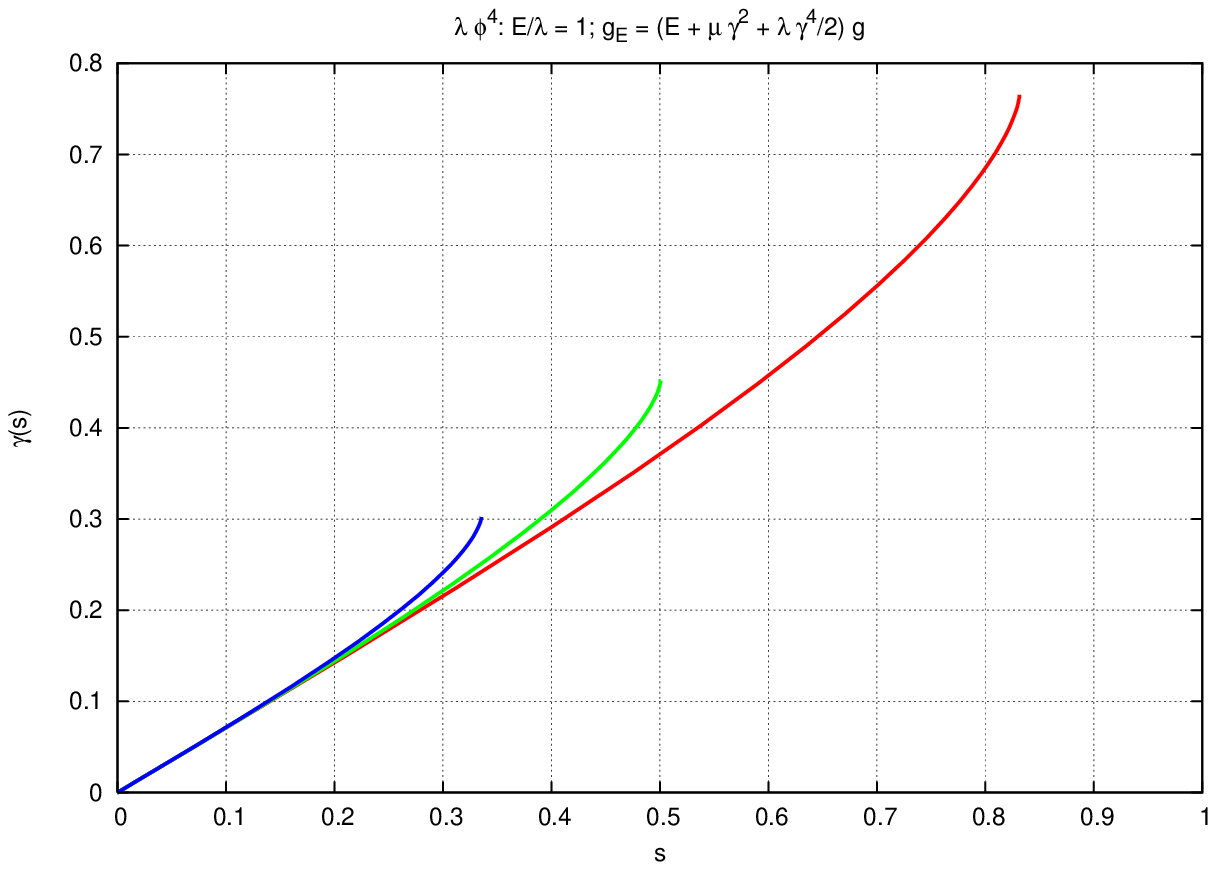} \hfill
\end{center}

We can clearly see that, for positive $\Delta$ (leftmost graph), we have a
smooth geodesic $\gamma(s)$ representing the symmetric phase. Then, when
$\Delta$ vanishes (middle graph), there is a clear change which delimits
these 2 different phases of the theory. Moreover, when $\Delta$ is negative
(rightmost graph), we have the third [broken-symmetric] phase of the theory
(which has a finite geodesic). Note that just as mentioned before, this
construction is made at every point in spacetime; therefore, the number of
solutions is 3 times infinity.
\subsection{The Landau-Ginzburg Functional}\label{subsec:lgf}
To start off, let us consider the case where the base manifold is a compact Riemann
surface $\Sigma$ equipped with a conformal metric and the vector bundle is a Hermitian
line bundle $\mathpzc{L}$ (\ie, with fiber $\mathbb{C}$ and a Hermitian metric
$\hm{\cdot}{\cdot}$ on the fibers).

The Landau-Ginzburg functional is defined for a section $\varphi$ and a unitary connection
$D_A = \ed\, + A$ of $\mathpzc{L}$ as ($\sigma\in\mathbb{R}$ is a real scalar),

\begin{equation*}
  L(\varphi,A) = \int_{\Sigma} |F_A|^2 + |D_A\, \varphi|^2 + \frac{1}{4}\, \bigl(\sigma -
    |\varphi|^2\bigr)^2 \; .
\end{equation*}

Thus, its Euler-Lagrange equations [of motion] are given by:
\begin{align*}
  D_A^{*}\, D_A\, \varphi &= \frac{1}{2}\, \bigl(\sigma - |\varphi|^2\bigr)\, \varphi\;;\\
  D_A^{*}\, F_A &=  -\re\hm{D_{A}\, \varphi}{\varphi} \; ;
\end{align*}
where, $D_A^{*}$ is the dual of $D_A$, \ie, $D_A^{*} = -*\, D_A\, * = -*\, (\ed\,
+ A)\, *$. Note that the equation for $F_A$ (the second one above) is linear 
in $A$. Since $D_A$ is a unitary connection, $A$ is a $\mathfrak{u}(1)$-valued
1-form. This Lie algebra (of the group $U(1)$) will sometimes be identified with
$i\, \mathbb{R}$ --- in other words, our $A$ corresponds to $-i\, A$ in the
standard physics literature (where $A$ is real-valued).

Before we go any further, some notational remarks are in order. We decompose the
space of 1-forms, $\Omega^1$, on $\Sigma$ as $\Omega^1 = \Omega^{1,0} \oplus
\Omega^{0,1}$, with $\Omega^{1,0}$ spanned by 1-forms of the type $dz$ and
$\Omega^{0,1}$ by 1-forms of the type $d\bar{z}$. Here $z = x + i\, y$ is a
local conformal parameter on $\Sigma$ and $\bar{z} = x - i\, y$. Therefore, $dz
= dx + i\, dy$, $d\bar{z} = dx - i\, dy$, $\partial_z = \tfrac{1}{2}\,
(\partial_x - i\, \partial_y)$ and $\partial_{\bar{z}} = \tfrac{1}{2}\,
(\partial_x + i\, \partial_y)$. Furthermore, if $\partial_x$ and $\partial_y$
are an orthonormal basis of the tangent space of $\Sigma$ at a given point, we
have that $\hm{dz}{dz} = 2$, $\hm{d\bar{z}}{d\bar{z}} = 2$ and
$\hm{dz}{d\bar{z}} = 0$. Given that the decomposition $\Omega^1 = \Omega^{1,0}
\oplus \Omega^{0,1}$ is orthogonal, we may also decompose $D_A$ accordingly:
$D_A = \partial_A + \bar{\partial}_A$, where $\partial_A\varphi \in
\Omega^{1,0}(\mathpzc{L})$ and $\bar{\partial}_A\varphi \in
\Omega^{0,1}(\mathpzc{L})$ for all sections $\varphi$ of $\mathpzc{L}$
(holomorphic, $\bar{\partial}_A\, f(z,\bar{z}) = 0 \Leftrightarrow
f(z,\bar{z}) = f(z)$, and anti-holomorphic, $\partial_A f(z,\bar{z}) = 0
\Leftrightarrow f(z,\bar{z}) = f(\bar{z})$, parts; \ie, the space of 1-forms
and the space of connections is decomposable into a direct sum of its
holomorphic and anti-holomorphic parts: $\partial_A = \partial + A^{1,0}$ and
$\bar{\partial}_A = \bar{\partial} + A^{0,1}$; while the exterior derivative is
given by $\ed\, = \partial + \bar{\partial}$). As expected, we have that
$\partial_A\, \partial_A = 0 = \bar{\partial}_A\, \bar{\partial}_A$ and $F_A =
-\bigl(\partial_A\, \bar{\partial}_A + \bar{\partial}_A\, \partial_A\bigr)$.

It is not difficult to show \cite{geomana} that:

\begin{align*}
    L(\varphi,A) &= \int_{\Sigma} |F_A|^2 + |D_A\, \varphi|^2 + \frac{1}{4}\, \bigl(\sigma -
      |\varphi|^2\bigr)^2 \; ; \\
    &= 2\, \pi\, \deg(\mathpzc{L}) + \int_{\Sigma} 2\, \bigl|\bar{\partial}_A\,
      \varphi\bigr|^2 + \Bigl(*(-i\, F_A) - \frac{1}{2}\, \bigl(\sigma -
      |\varphi|^2\bigr)\Bigr)^2 \; ; \\
    \text{where}\quad \deg(\mathpzc{L}) &= c_1(\mathpzc{L}) = \frac{i}{2\, \pi}\, \tr(F_A)\; ;
\end{align*}
\ie, the degree of the line bundle is given by the 1st Chern class.

Therefore, a very useful consequence from the above is that if $\deg(\mathpzc{L})
\geqslant 0$ the lowest possible value for $L(\varphi,A)$ is realized if $\varphi$ and $A$
satisfy

\begin{align}
  \label{eq:lg1}
  \bar{\partial}_A \, \varphi &= 0 \; ;\\
  \label{eq:lg2}
  *(i\, F_A) &= \frac{1}{2}\, \bigl(\sigma - |\varphi|^2\bigr) \; .
\end{align}

These equations are just the expression of the self-duality of the
Landau-Ginzburg functional. If $\deg(\mathpzc{L}) < 0$, then these equations
cannot have any solution, thus one has to consider the self-duality equations
arising from the Landau-Ginzburg functional where the term $+2\, \pi\,
\deg(\mathpzc{L})$ is substituted by $\boldsymbol{-}2\, \pi\,
\deg(\mathpzc{L})$. Therefore, without loss of generality, we shall assume
$\deg(\mathpzc{L}) \geqslant 0$. A necessary condition for the
solvability of \eqref{eq:lg2} is that,

\begin{align}
  \nonumber
  2\, \pi\, \deg(\mathpzc{L}) = \int i\, F_A &= \frac{1}{2}\, \int_{\Sigma} \bigl(\sigma -
    |\varphi|^2\bigr) \leqslant \frac{\sigma}{2}\, \text{Area}(\Sigma)\; ; \\
  \label{eq:lgentropy}
  \therefore\; \sigma &\geqslant \frac{4\, \pi \deg(\mathpzc{L})}{\text{Area}(\Sigma)} \; ;
\end{align}
and the equality only occurs if, and only if, $\varphi \equiv 0$.

The following result is useful when studying the solutions of the above
functionals \cite{geomana}: \emph{Let $\Sigma$ be a compact Riemann surface with
  a conformal   metric and $\mathpzc{L}$ as before. For any solution of
  \eqref{eq:lg1}, we have that $|\varphi| \leqslant \sigma$ on $\Sigma$.} That
is, this maximum principle states that the amplitude of the field cannot exceed
the height of the potential (see the first plot below).

Let us now construct the Jacobi metric for this potential and find its possible
geodesics:
\begin{align*}
  \gE &= 2\, \big(E - V_{\sigma}(\gamma)\big)\, \g \; ; \\
  &= 2\, \bigg(E + \frac{1}{4}\,\big(\sigma - |\gamma|^2\big)^2\bigg)\, \g \; .
\end{align*}

In a complete analogy to what was previously done, we consider the $P(\gamma) =
E - V(\gamma)$ polynomial. However, in order to make this analysis clearer, let
us, in fact, use a slightly different polynomial given by $P^{\prime}(\gamma) =
4\, \bigl(P(\gamma) - E\bigr)$. It is straightforward to see that $P'(\gamma) =
\bigl(|\gamma|^2 - \sigma\bigr)\, \bigl(|\gamma|^2 - \sigma\bigr)$, which means
that $\sigma$ is the only root of $P'(\gamma)$, with double multiplicity (the 2
roots coalesce into 1).

This situation implies that the discriminant of $P'(\gamma)$ vanishes, \ie,
$\Delta = 0$, and the different phases of the theory are labelled by $\sigma =
0$ and $\sigma > 0$.

The plots below have, respectively, the following
values for $\sigma$: $0.0$, $3.0$, $5.0$, $7.0$, $11.0$ and $31.0$. As they
show, when $\sigma = 0$ we have $c_1(\mathpzc{L}) = \deg(\mathpzc{L}) =
\frac{i}{2\, \pi}\, \tr(F_A) = 0$ and its geodesic has a clear character which
is quite different otherwise:

\begin{center}
  \includegraphics[scale=0.4]{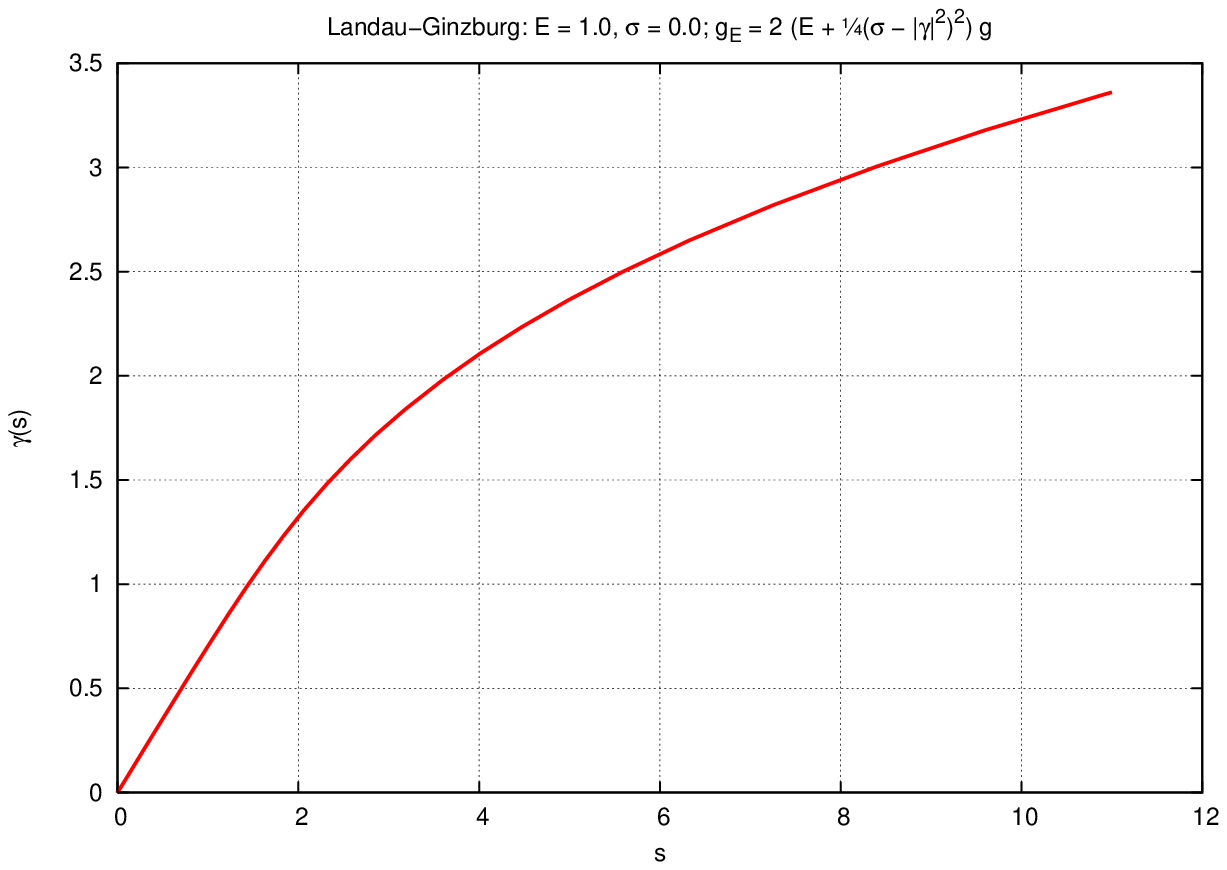} \hfill
  \includegraphics[scale=0.4]{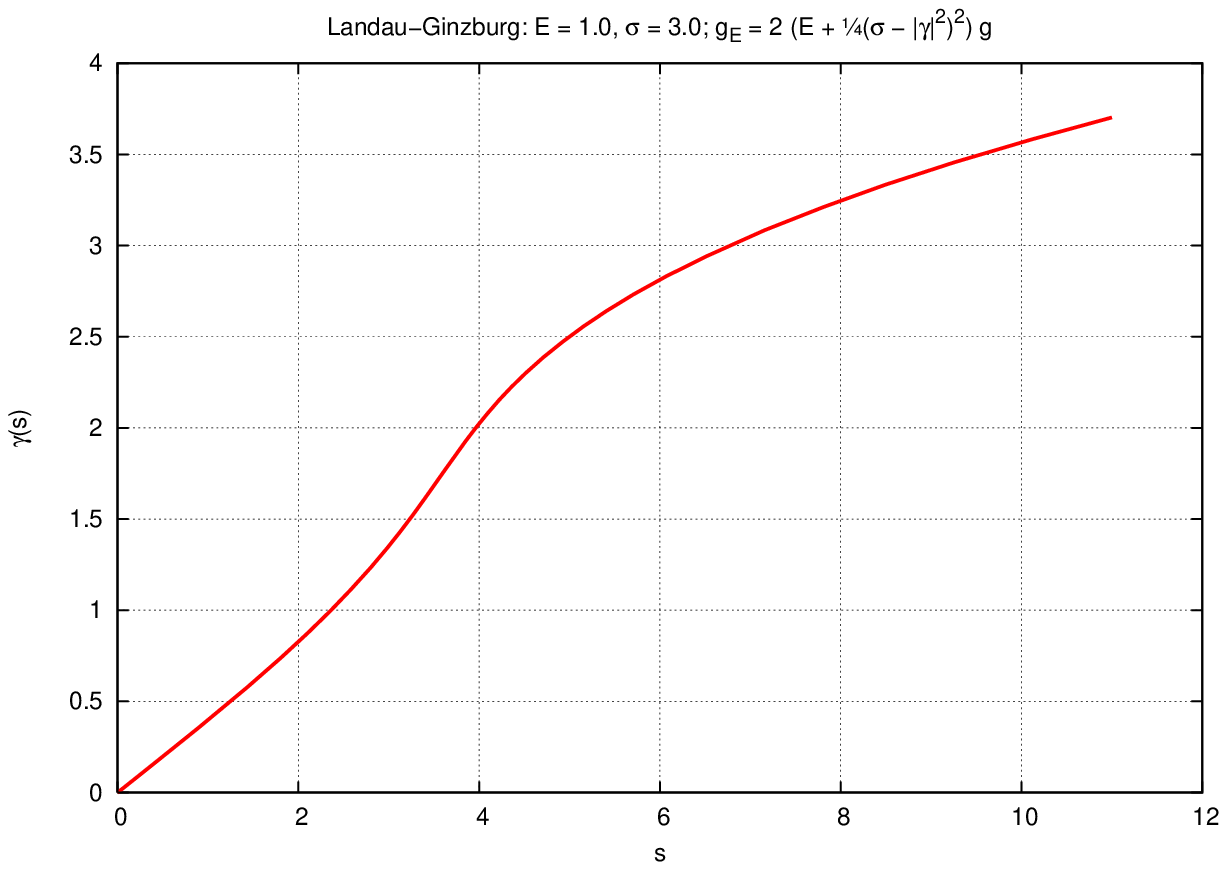} \hfill
  \includegraphics[scale=0.4]{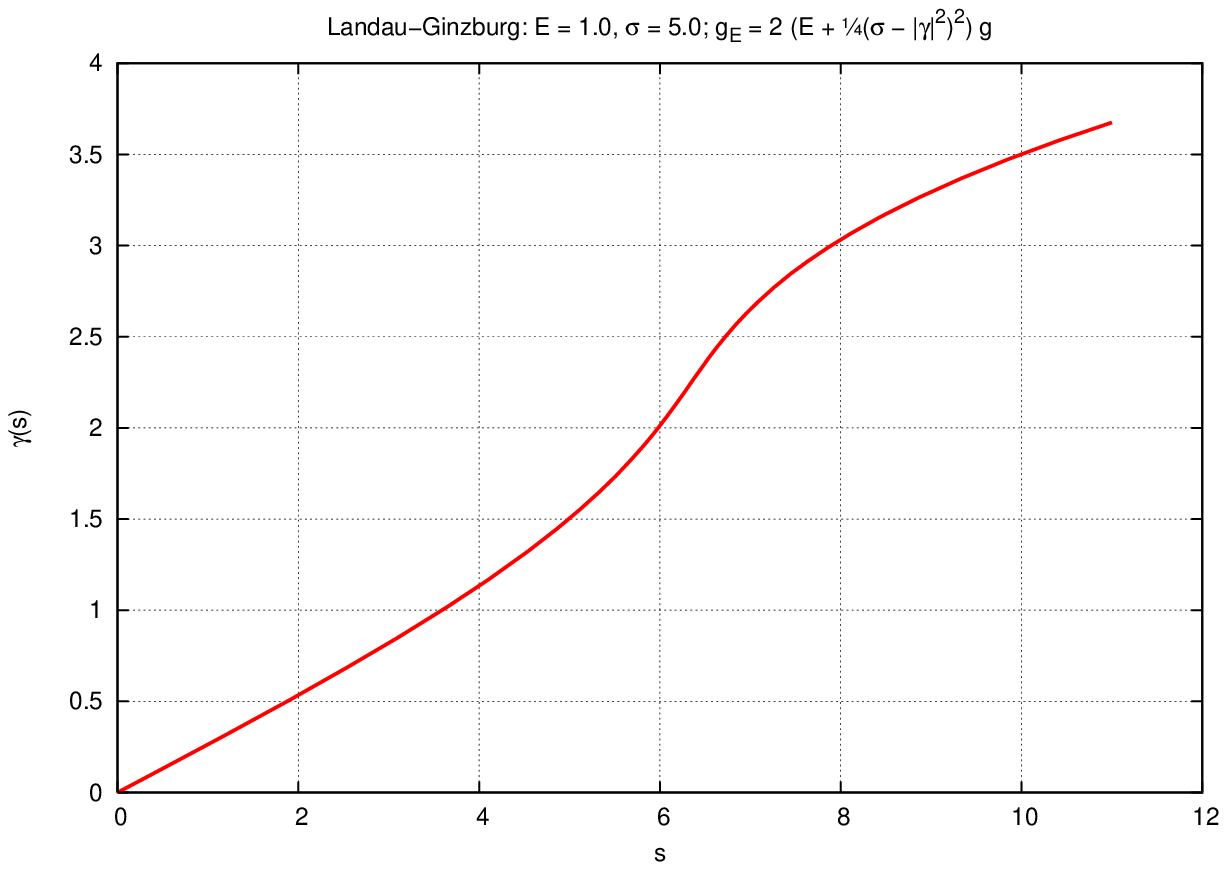} \hfill
  \includegraphics[scale=0.4]{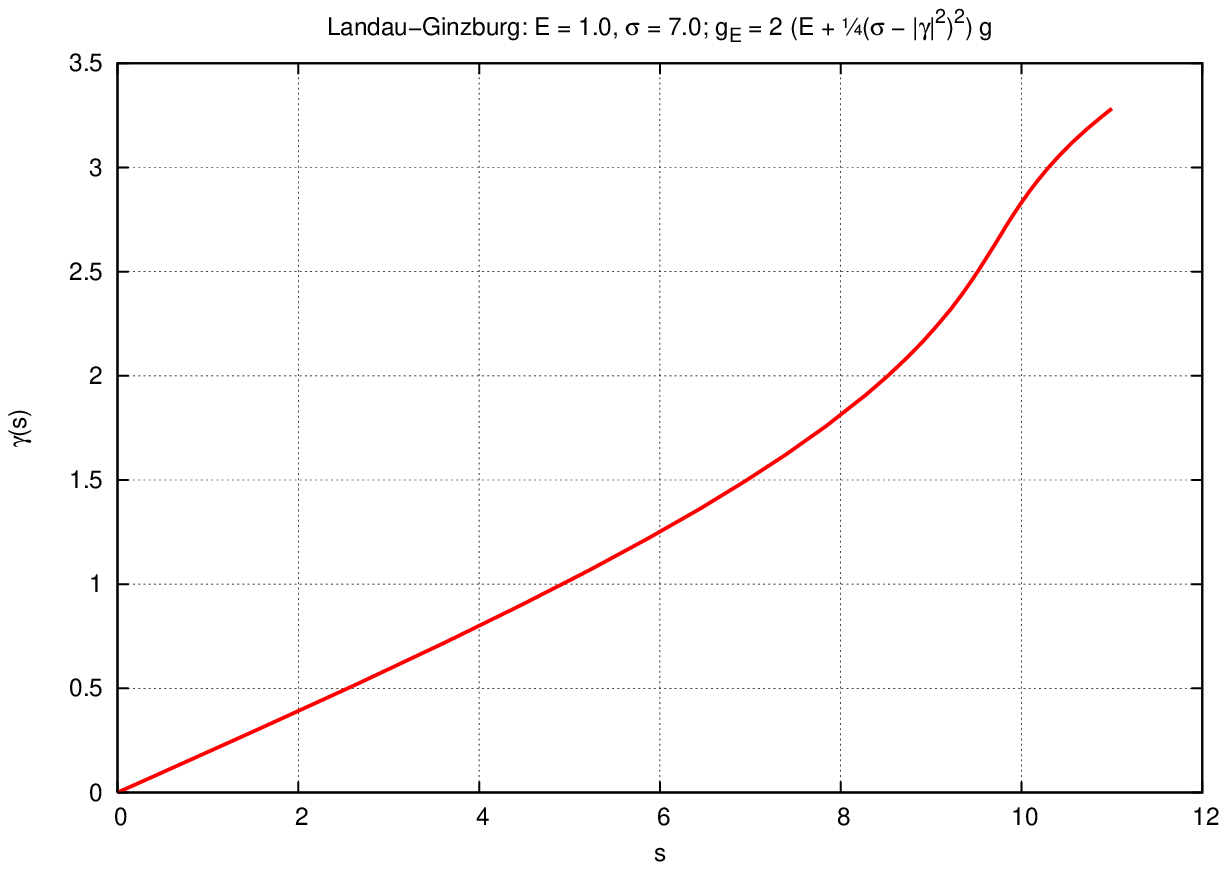} \hfill
  \includegraphics[scale=0.4]{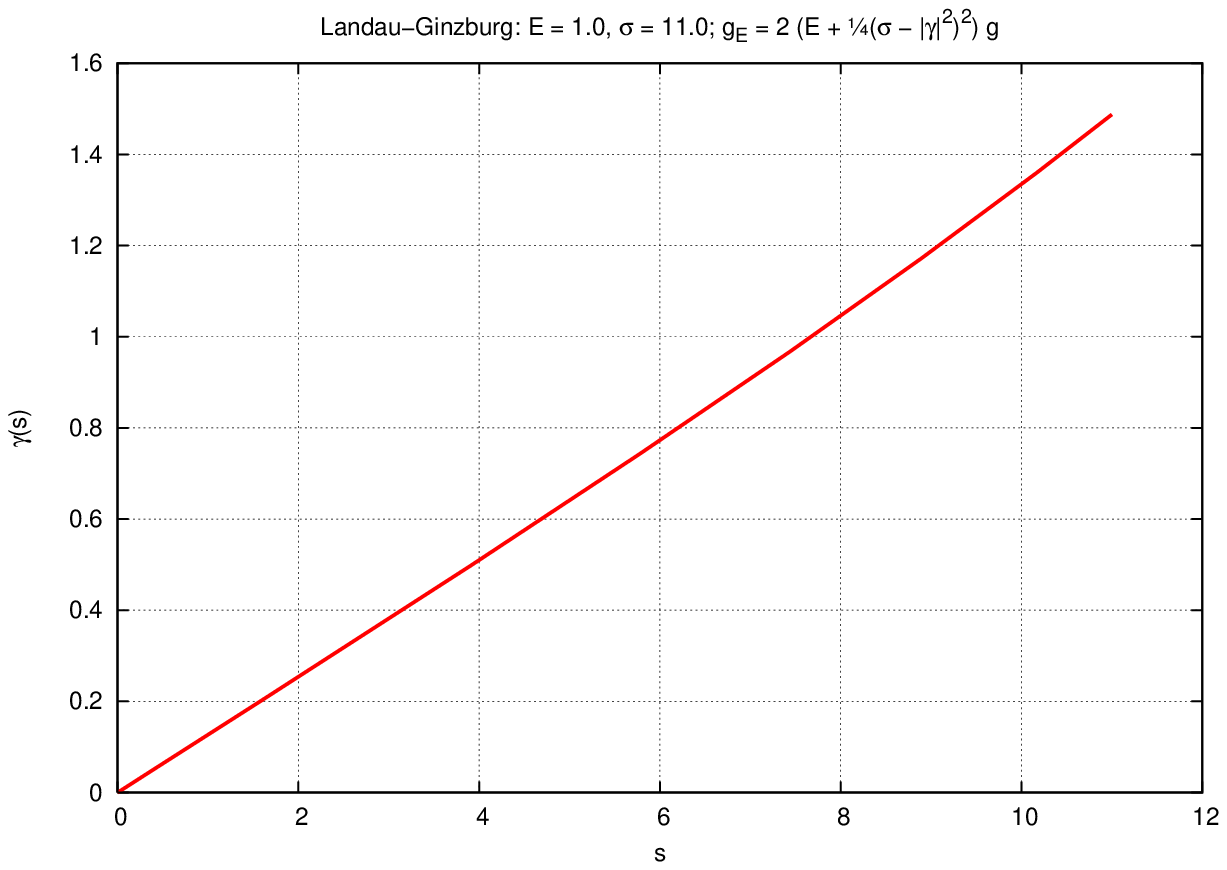} \hfill
  \includegraphics[scale=0.4]{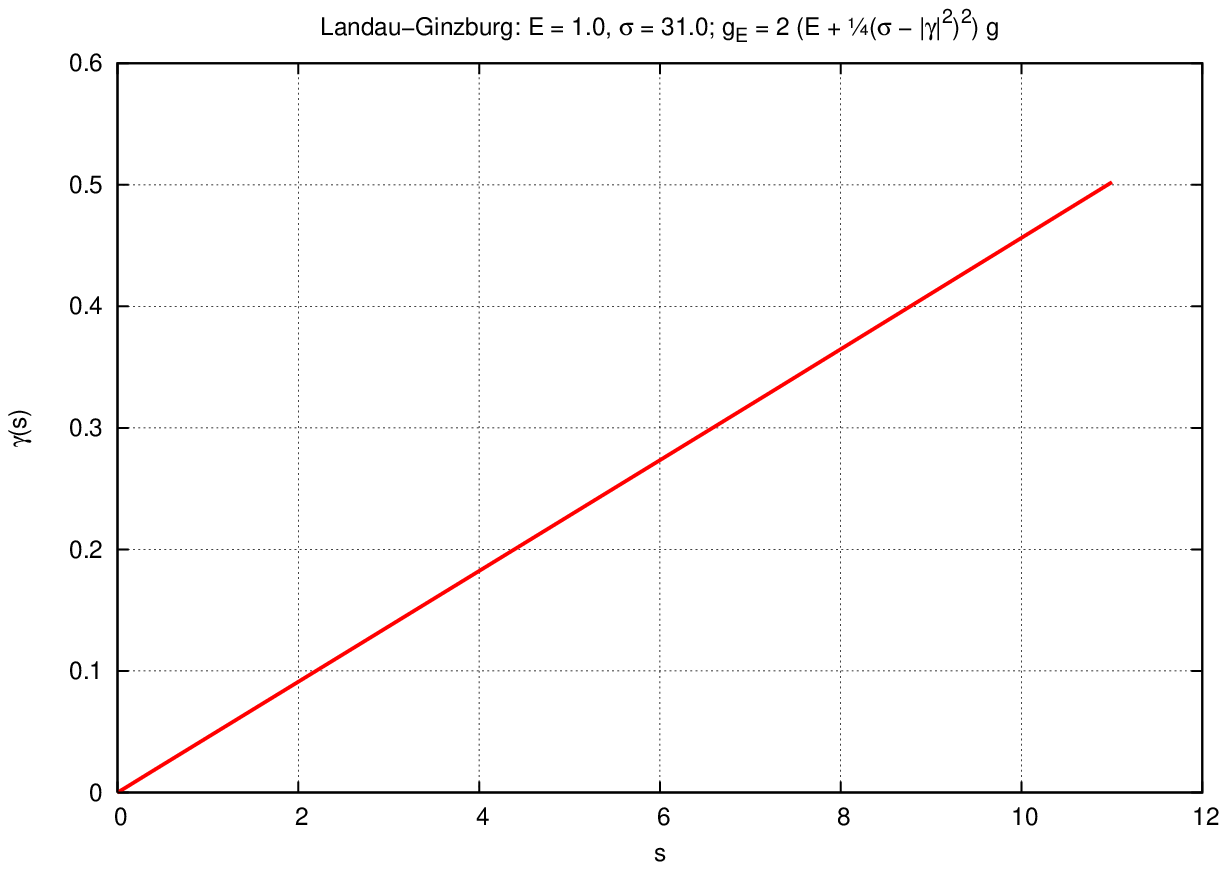} \hfill
\end{center}
%
\subsection{The Seiberg-Witten Functional}\label{subsec:swf}
In this case, the base manifold, $\mathpzc{M}$, is a compact, oriented,
4-dimensional Riemannian manifold endowed with a spin${}^c$ structure, \ie, a
spin${}^c$ manifold. The determinant line of this spin${}^c$ structure will be
denoted by $\mathpzc{L}$ and the Dirac operator determined by a unitary
connection $A$ on $\mathpzc{L}$ will be denoted by $\mathcal{D}_A$. Recalling
the half spin bundle $\mathcal{S}^{\pm}$ defined by the spin${}^c$ structure, we
see that $\mathcal{D}_A$ maps sections of $\mathcal{S}^{\pm}$ into sections of
$\mathcal{S}^{\mp}$ \cite{seibergwitten,donaldson,liviu,geomana}.

In this fashion the Seiberg-Witten functional for a unitary connection $A$ on
$\mathpzc{L}$ and a section $\varphi$ of $\mathcal{S}^+$ is given by,

\begin{equation}
  \label{eq:swf}
  SW[\varphi,A] = \int_{\mathpzc{M}} |\nabla_A \varphi|^2 + |F^+_A|^2 + \frac{R}{4}\,
    |\varphi|^2 + \frac{1}{8}\, |\varphi|^4 \; ;
\end{equation}
where $\nabla_A$ is the spin${}^c$ connection induced by $A$ and the Levi-Civita
connection of $\mathpzc{M}$, $F_A^+$ is the self-dual part of the curvature of $A$ and $R$
is the scalar curvature of $\mathpzc{M}$. Its Euler-Lagrange equations are given by,

\begin{align}
  \label{eq:sw1}
  \nabla_A^*\, \nabla_A\, \varphi &= -\biggl(\frac{R}{4} + \frac{1}{4}\, |\varphi|^2
    \biggr)\, \varphi \; ; \\
  \label{eq:sw2}
  \ed{}^*\, F_A^+ &= -\re\hm{\nabla_{A}\, \varphi}{\varphi} \; .
\end{align}

Using a spin frame, it is not difficult \cite{geomana} to show that the Seiberg-Witten
functional can be written in the following form:

\begin{equation*}
  SW[\varphi,A] = \int_{\mathpzc{M}} |\mathcal{D}_A \varphi|^2 + \biggl|F_A^+ -
    \frac{1}{4}\, \hm{e_j\cdot e_k\cdot \varphi}{\varphi}\, e^j \wedge e^k\biggr|^2 \; ;
\end{equation*}
where $e^j$ are 1-forms dual to the tangent vectors $e_j$, $e^j(e_k) = \delta^j_k$;
$j,k=1,\dotsc,4$. As a corollary of the above, the lowest possible value of the
Seiberg-Witten functional is achieved if $\varphi$ and $A$ are solutions of the
\emph{Seiberg-Witten equations}:

\begin{align}
  \label{eq:sw3}
  \mathcal{D}_A \varphi &= 0 \; ;\\
  \label{eq:sw4}
  F_A^+ &= \frac{1}{4}\, \hm{e_j\cdot e_k\cdot \varphi}{\varphi}\, e^j \wedge e^k \; .
\end{align}

Thus, self-duality is at work yet again: the absolute minima of the Seiberg-Witten
functional satisfy not only the second order equations \eqref{eq:sw1} and \eqref{eq:sw2},
but also the first order Seiberg-Witten equations \eqref{eq:sw3} and \eqref{eq:sw4}.

Although our discussion of the Seiberg-Witten functional, so far, has mirrored our
discussion of the Landau-Ginzburg one, the parameter $\sigma$ on the latter has had no
analogue in the former. This can be accomplished with the introduction of a 2-form $\mu$
and the consideration of the perturbed functional,

\begin{align*}
  SW_{\mu}[\varphi,A] &= \int_{\mathpzc{M}} |\mathcal{D}_A \varphi|^2 + \biggl|F_A^+ -
    \frac{1}{4}\, \hm{e_j\cdot e_k\cdot \varphi}{\varphi}\, e^j \wedge e^k +
    \mu\biggr|^2 \; ; \\
  &= \int_{\mathpzc{M}} |\nabla_A \varphi|^2 + |F^+_A|^2 + \frac{R}{4}\,
    |\varphi|^2 + \biggl|\mu - \frac{1}{4}\, \hm{e_j\cdot e_k\cdot \varphi}{\varphi}\, e^j
    \wedge e^k\biggr|^2 + 2\, \hm{F_A^+}{\mu} \; .
\end{align*}
Their corresponding first order equations of motion are,

\begin{align*}
  \mathcal{D}_A \varphi &= 0 \; \\
  F^+_A &= \frac{1}{4}\, \hm{e_j\cdot e_k\cdot \varphi}{\varphi}\, e^j\wedge e^k - \mu\;.
\end{align*}

If we assume that $\mu$ is closed and self-dual, then we see that $\hm{F_A}{\mu} =
\hm{F_A^+}{\mu}$, once $\hm{F_A^-}{\mu} = 0$ due to the orthogonality between
anti-self-dual and self-dual forms. Thus, given that $F_A$ represents the first Chern
class $c_1(\mathpzc{L})$ of the line bundle $\mathpzc{L}$, and we assumed $\mu$ to be
closed (hence it represents a cohomology class $[\mu]$), the integral

\begin{equation*}
  \int_{\mathpzc{M}} \hm{F_A}{\mu} \; ,
\end{equation*}
does not depend on the connection $A$, thus representing a topological invariant, denoted
by $\bigl(c_1(\mathpzc{L})\wedge[\mu]\bigr)[\mathpzc{M}]$.

Just as before, we also have a maximum principle: \emph{For any solution $\varphi$ of
  \eqref{eq:sw1} --- in particular, for any solution of \eqref{eq:sw3} --- on a compact
  4-dimensional Riemannian manifold, we have that,}

\begin{equation*}
  \max_{\mathpzc{M}} |\varphi|^2 \leqslant \max_{x\in \mathpzc{M}} (-R(x), 0)\; .
\end{equation*}

As a direct consequence of this, if the compact, oriented, Riemannian Spin${}^c$ manifold
$\mathpzc{M}$ has nonnegative scalar curvature, the only possible solution of the
Seiberg-Witten equations is,

\begin{equation*}
  \varphi\equiv 0\; ; \quad F_A^+ \equiv 0 \; .
\end{equation*}

The Jacobi metric for this Seiberg-Witten model is given by

\begin{equation*}
  \gE = 2\, \biggl(E + \frac{R}{4}\, |\gamma|^2 + \frac{1}{8}\, |\gamma|^4
    \biggr)\, \g \; ;
\end{equation*}
for a geodesic $\gamma$; and, just like before, although $P(\gamma) = E -
V(\gamma) = E + R\, |\gamma|^2/4 + |\gamma|^4/8$, let us consider $P'(\gamma) =
8\, \bigl(P(\gamma) - E\bigr) = \bigl(|\gamma|^2 - 0\bigr)\, \bigl(|\gamma|^2 +
2\, R\bigr)$. It is then clear that when $R = 0$ the discriminant vanishes
$(\Delta = 0)$ and the 2 roots merge into 1, what constitutes one of the phases
of the theory. When $R \neq 0$, the discriminant is either $\Delta > 0$ $(R >
0)$ or $\Delta < 0$ $(R < 0)$, which accounts for the other 2 phases of the
theory.

The plots below were obtained with the choice of $E = 1.0$ and $R$ respectively
equal to $0.0$, $3.0$, $7.0$, $-3.0$, $-5.0$, $-7.0$.

\begin{center}
  \includegraphics[scale=0.4]{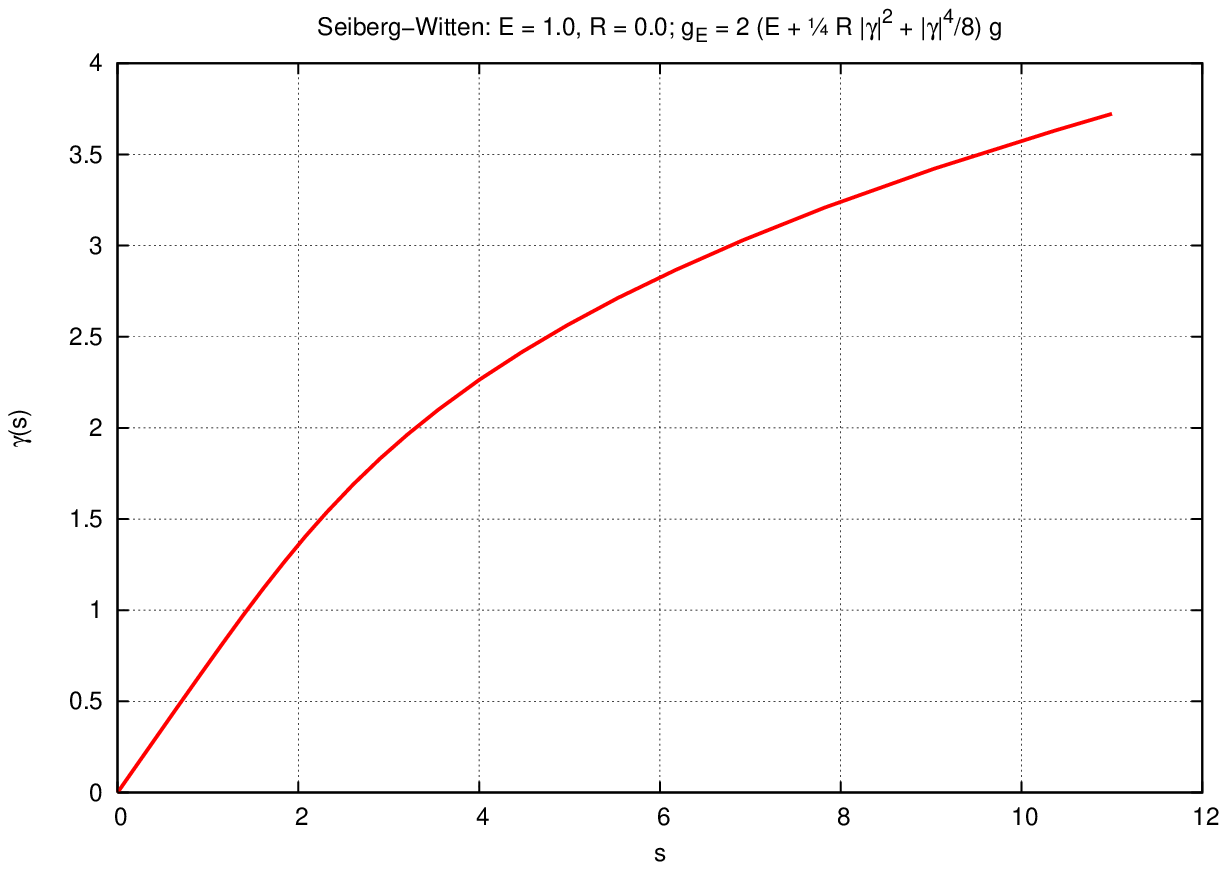} \hfill
  \includegraphics[scale=0.4]{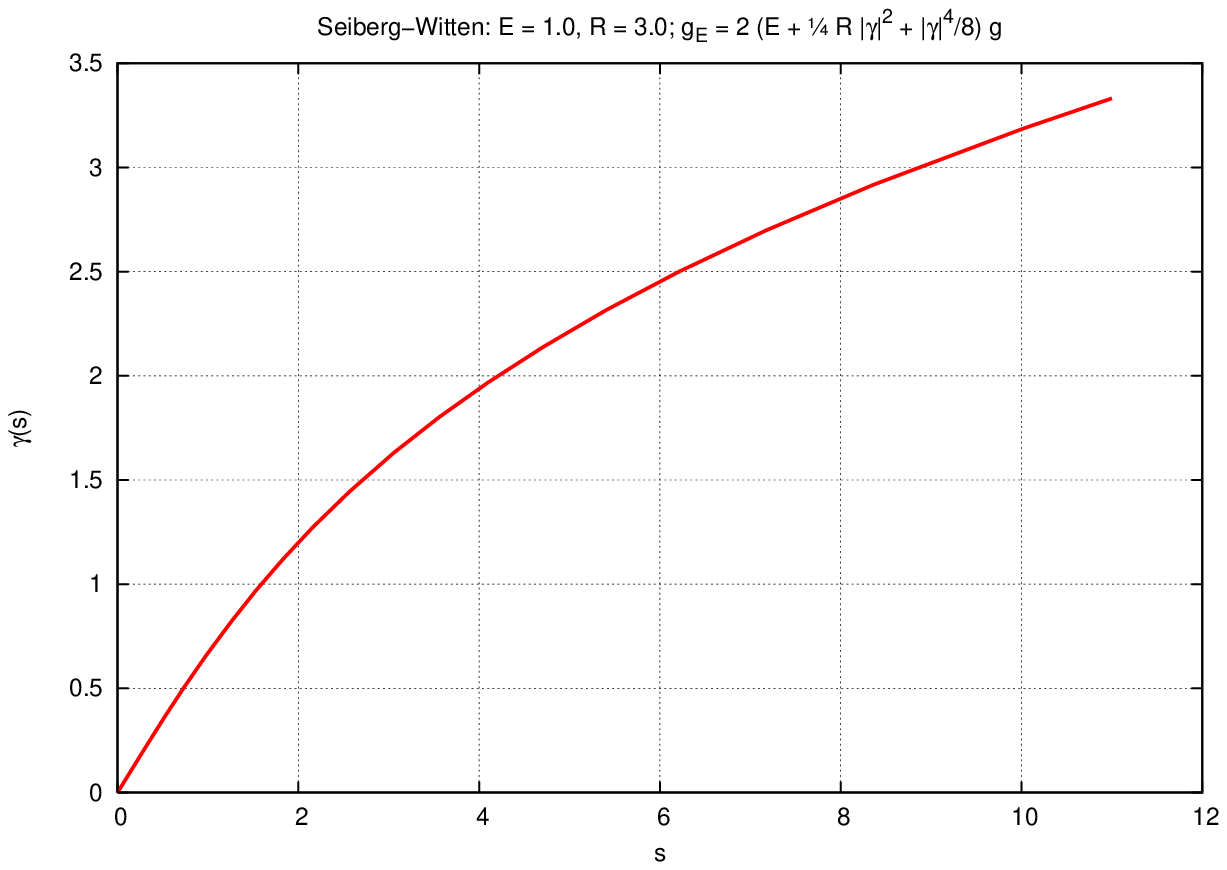} \hfill
  \includegraphics[scale=0.4]{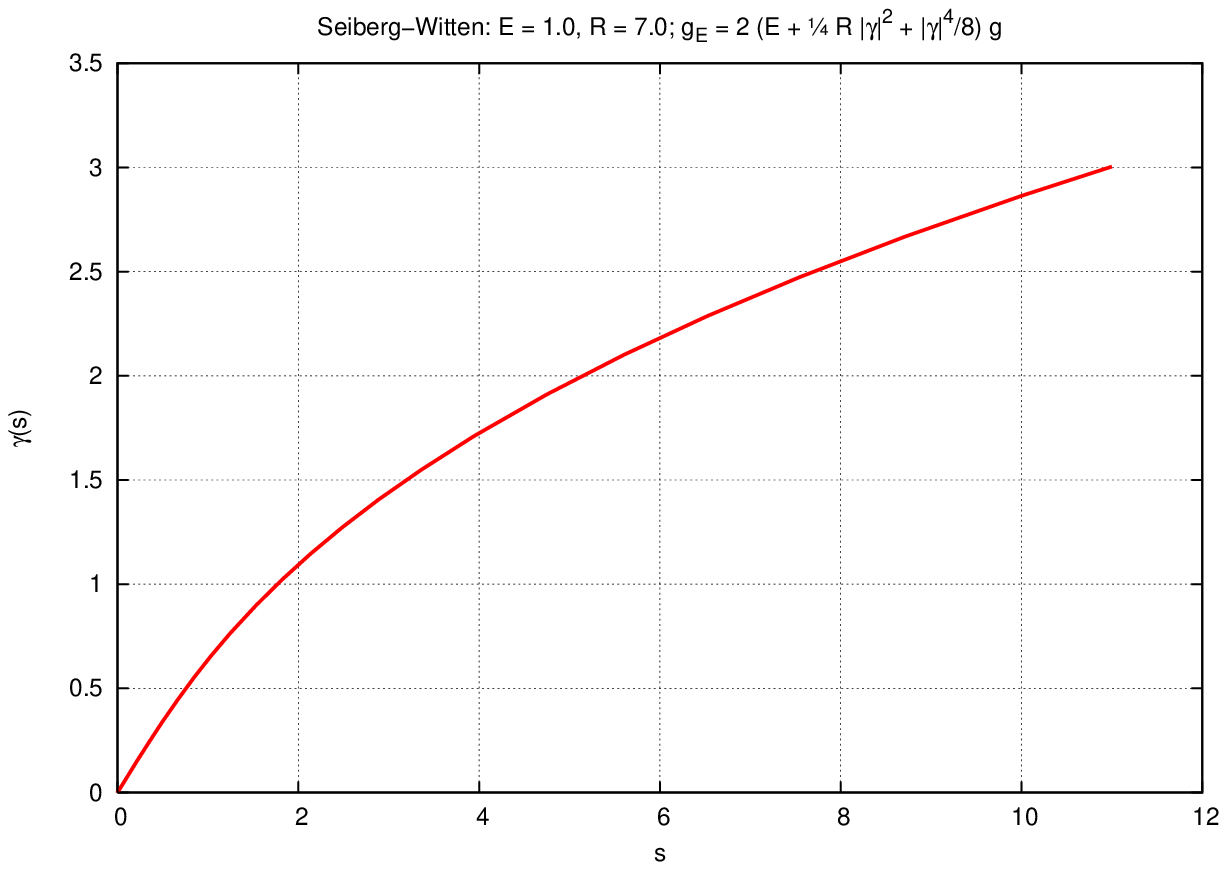} \hfill
  \includegraphics[scale=0.4]{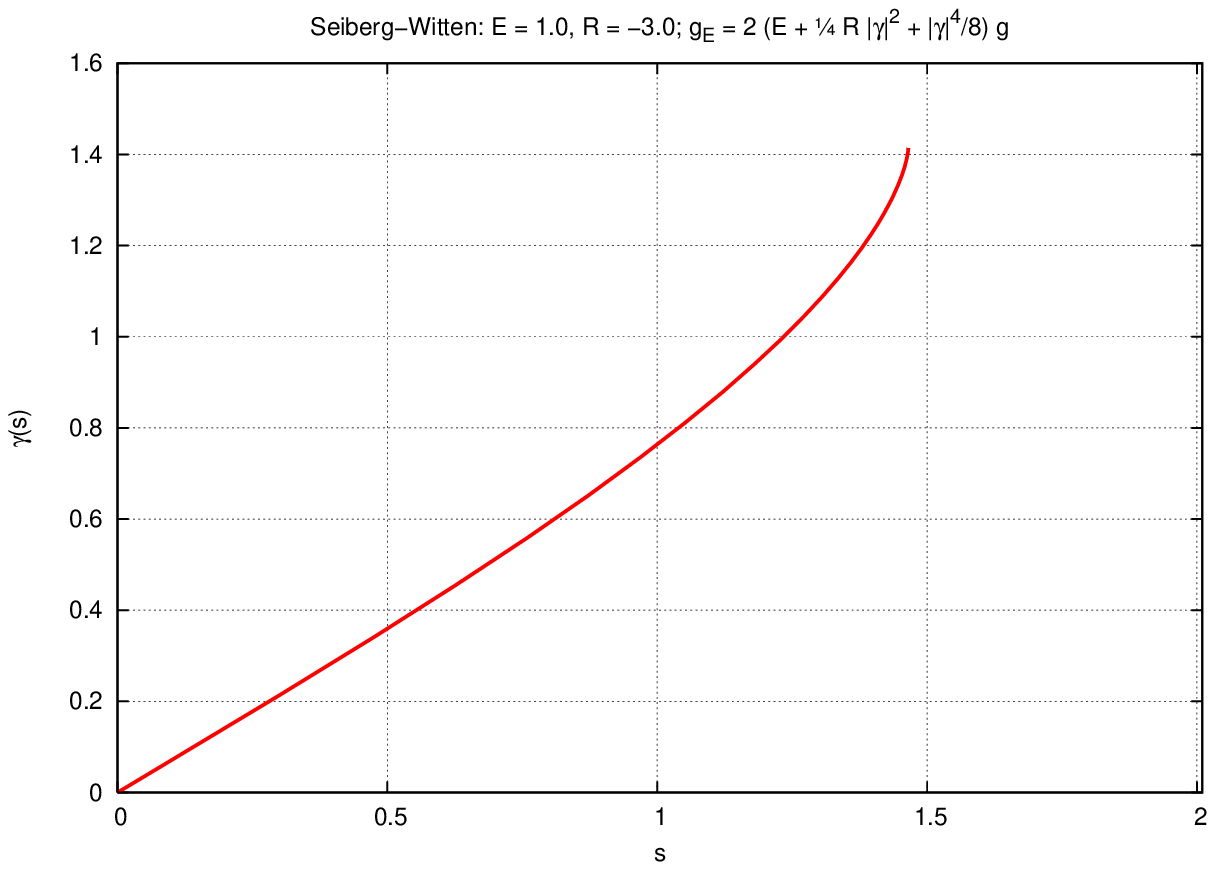} \hfill
  \includegraphics[scale=0.4]{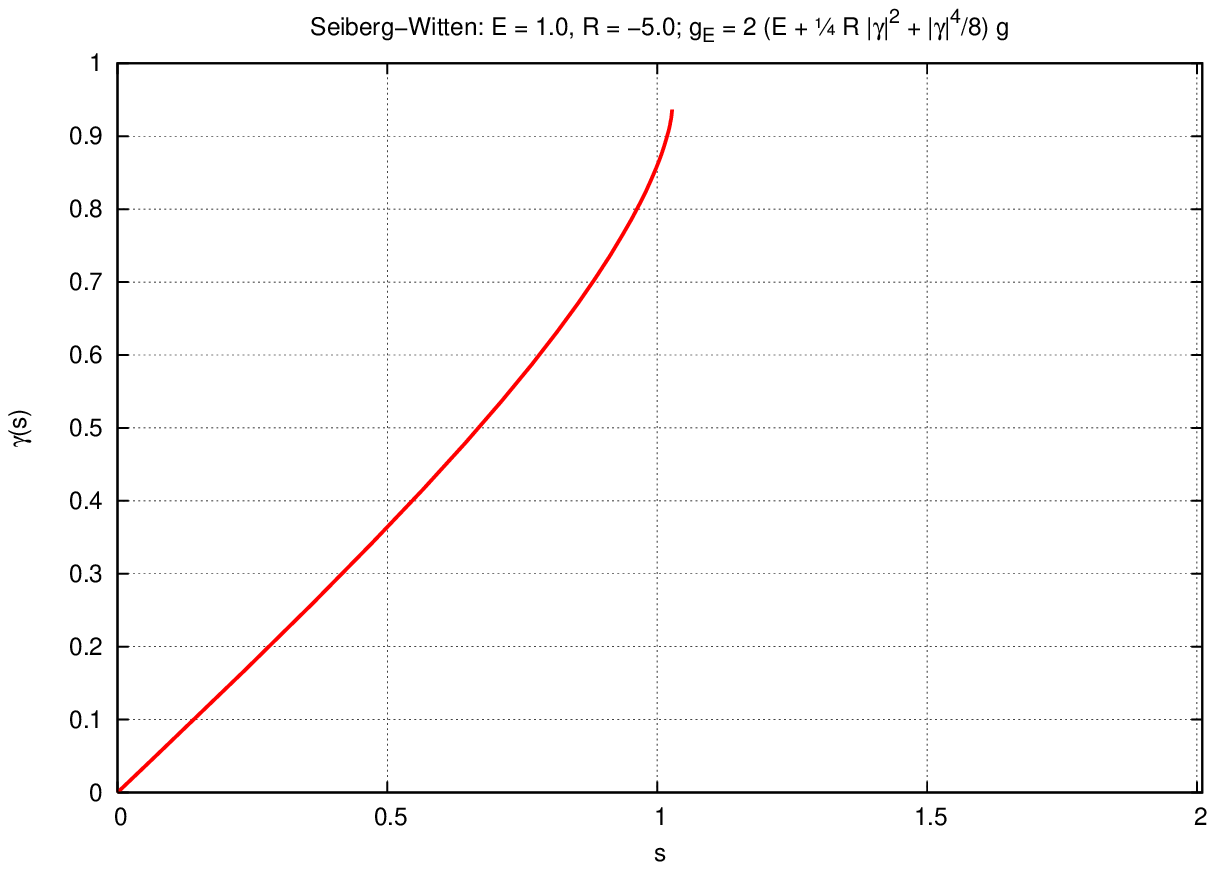} \hfill
  \includegraphics[scale=0.4]{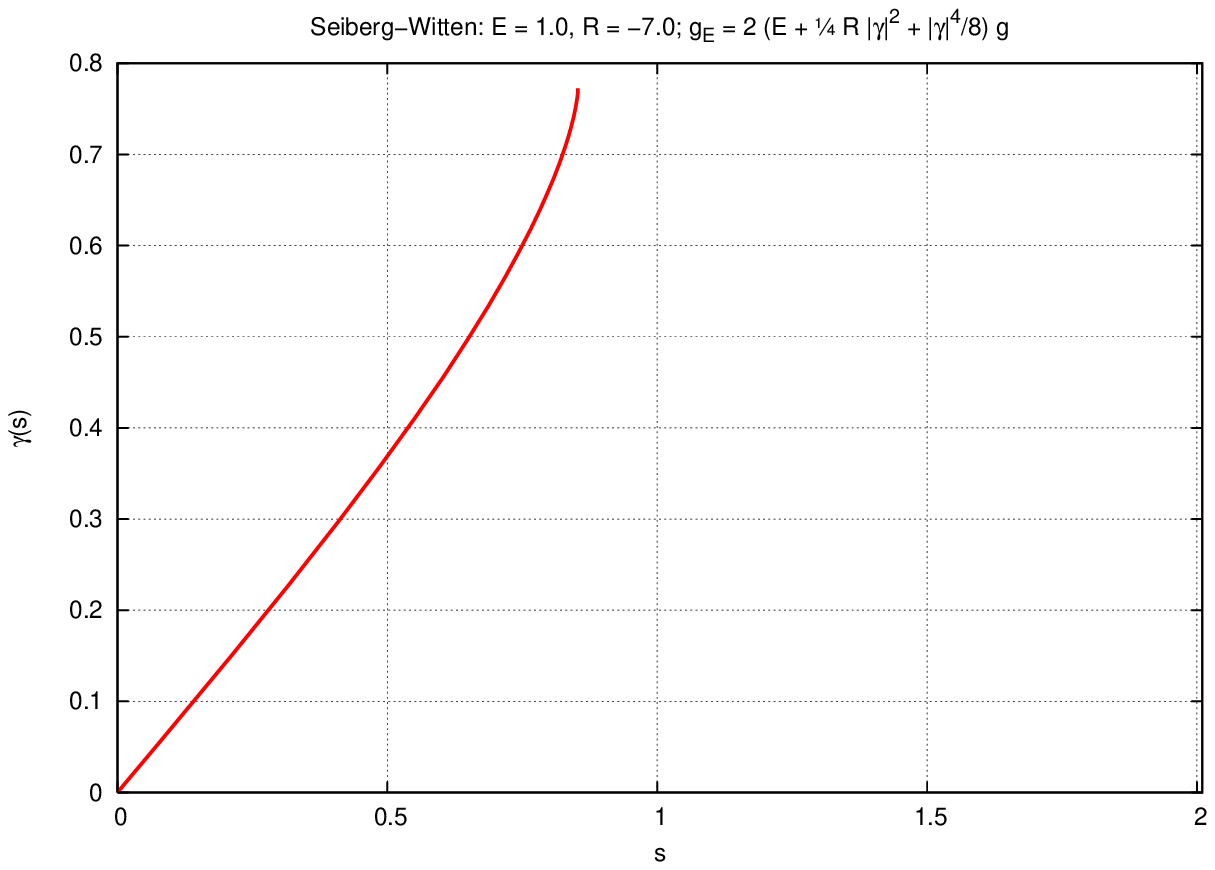} \hfill
\end{center}
\section{Conclusions} \label{sec:cc}
Historically, the conformal transformation \eqref{eq:jacobimetric} was used by
Jacobi with a twofold purpose: unify Hamilton's (mechanics) and Fermat's
(geometrical optics) action principles into a single framework, and turn every
Hamiltonian flow into a geodesic one \cite{frankel,drg}. The extension
of the Jacobi metric from the Riemannian setting to the Lorentzian one was done
in \cite{marek}.

Note that this construction extends to field theory, once the
space of solutions of the field equations is metrizable: fields satisfying the
equations of motion are particular sections of a principal fibre bundle, $P$,
that belong to $\ker\bigl\{\Box + V'(\phi)\bigr\}$; in this sense, if $P$ is
endowed with a metric, its fibers inherit an induced metric and our construction
of \eqref{eq:jacobimetric} is nothing but a conformal transformation of such a
metric.

However, rather than dealing with the phase space of the [quantum] field theory
in question, we restrain ourselves to studying its geodesic structure.

In this sense, we have shown that this geodesic structure is quite rich and its
different parameters label different phases of the theory. Moreover, the
technique works for both, global (subsection \ref{subsec:lp4}) and local (gauge)
symmetries (subsections \ref{subsec:lgf} and \ref{subsec:swf}).

As mentioned in the introduction, this method can be used in order to replace
the Coleman-De Luccia mechanism in string theoretical models. The way to do this
is to use Jacobi's metric in order to translate between the spacetime geometry
and a potential $V$ for a Lagrangian describing the particular model in
question. In this sense, the theory at hand can be thought of as a QFT, and we
can repeat the analysis outlined in this paper looking for the possible
[topologically inequivalent] phases of the string theoretical model being
treated.

Therefore, in this string theoretical setting, the picture that emerges is that
quantum effects (in dynamically breaking the symmetry) would be responsible for
the generation of topologically inequivalent ``baby universes''. Thus, there is
no need for the use of a Coleman-De Luccia type of mechanism in order to
generate bubbles of baby universes.

On a forthcoming work, we want to take this construction further, elucidating
its relation to Morse Theory and Deformation and Geometric Quantization, also
making clearer the connection with the Renormalization Group.
\section*{Acknowledgements}
The authors would like to thank C. Pehlevan for discussions and useful
conversations. This work is supported in part by funds provided by the US
Department of Energy (\textsf{DoE}) under \textsf{DE-FG02-91ER40688-TaskD}.
\end{document}